\documentclass[aps,12pts,pre,amsmath,amssymb,preprint,showkeys,showpacs]{revtex4-1}
\usepackage{graphicx}
\usepackage{bm}
\usepackage{mathptmx}
\usepackage{epstopdf}
\usepackage{booktabs}
\usepackage{multirow}
\usepackage{hhline}
\usepackage{dcolumn}
\usepackage{subfigure}
\usepackage{amsmath}
\usepackage{float}
\usepackage{chemformula}
\usepackage[version=3]{mhchem}
\usepackage{chemfig}

\begin{document}
\title{Uncovering the puzzle of complex magnetism in Fe$_{16}$N$_2$: a first-principles based study}
\author{Satadeep Bhattacharjee$^{1}$ and Seung-Cheol Lee$^{1,2}$}
\address{$^{1}$Indo-Korea Science and Technology Center (IKST), Bangalore, India \\
         $^{2}$Electronic Materials Research Center, Korea Institute of Science $\&$ Technology, Korea}

\begin{abstract}The electronic structure and magnetic exchange interactions in pure and V-doped Fe$_{16}$N$_2$ are studied within the framework of density functional theory. The Curie temperatures were obtained with both mean field approximation (MFA) as well as Monte Carlo (MC) calculations. The Curie temperature (T$_C$) for pure Fe$_{16}$N$_2$ obtained within MFA are significantly larger than the experimental value, suggesting the importance of thermal fluctuations in these systems, and has a resemblance of a lower dimensional spin system. We also briefly discuss about the various possible factors which may lead to a large magnetic moment in this material. The calculated magnetic susceptibility at zero field shows sharp peak at T=T$_C$ which resemble a local moment system. From the nature of exchange interactions we try to figure out the nature of the Fe-sites which might contain localized d-states.  Finally, we point out that Fe$_{16}$N$_2$ can also act as a good spin injector for the III-V semiconductors in addition to its well promised application as permanent magnet since it has a very high spin polarization (larger compared to elemental ferromagnets) as well as quite smaller lattice mismatch (compared to half-metallic Heusler alloys) with the conventional III-V semiconductors such as GaAs or InGaAs. We further demonstrate this through our calculations for Fe$_{16}$N$_2$(001)/InGaAs(001) heterostructures which shows the non-negligible spin polarization in the semiconductor (InGaAs) region implying a long spin diffusion length.

\end{abstract}

\maketitle

\section{Introduction}
Permanent magnets, which do not contain any rare-earths or platinum are a subject of recent attention due to their importance in the field of energy-saving technologies for the next generation of electrical devices. $\alpha^{''}$-Fe$_{16}$N$_2$, a martensite has been studied over years, as there were claims of giant magnetic saturation in this compound by several groups~\cite{giant1,giant2,giant3}. Historically, It was  experimentally first prepared by K. H Jack \cite{jack} while low temperature annealing of $\alpha^{'}$-FeN. It appeared as a metastable phase, which survived up to about 500K above which it was decomposed to $\alpha$-Fe and Fe$_4$N.

Existence  of giant saturation moment in $\alpha^{''}$-Fe$_{16}$N$_2$ was first reported by Kim \textit{et al.}~\cite{giant1} and once latter confirmed by Sugita \textit{et al.}~\cite{giant2} in single crystal Fe$_{16}$N$_2$ grown epitaxially on InGaAs. Since then, there have been several investigations exploring the validation of giant saturation moment in this material. However, these investigations resulted more controversies than bringing the issue to a firm conclusion since some of these studies confirmed the existence of giant moment~\cite{confirmed1,confirmed2} while the other were not able to reproduce such results~\cite{failed}.
 
\par
Research on Fe$_{16}$N$_2$ has therefore always faced two challenges: (1) to stabilize the material at higher temperature, so that it does not decompose to $\alpha$-Fe and Fe$_4$N, (2) to understand the complex magnetic behaviour of it which made the material as "\textit{40 year old mystery}"~\cite {myst} among the magnetic materials. In a recent work, we have shown that~\cite{KM}, Fe$_{16}$N$_2$ can be stabilized via Vanadium (V) doping. Recently Ke \textit{et al.}~\cite{kirill} have studied the exchange interaction and ferromagnetic transition temperature for Fe$_{16}$N$_2$, Co and Ti doped Fe$_{16}$N$_2$ using first-principle based method. They obtained the Curie temperature (T$_C$) using both mean field approximation (MFA) as well as random phase approximation (RPA). Direct measurement of T$_C$ of Fe$_{16}$N$_2$ is limited due to the decomposition of the material into Fe$_4$ and Fe at temperature  above 500 C~\cite{jack,giant1}. The theoretically obtained Curie temperatures by Ke \textit{et al.} were almost three times higher than one which was obtained by Sugita \textit{et al.} through extrapolating the experimental data obtained at lower temperature. 

                             Understanding the magnetism and the exchange mechanism in Fe$_{16}$N$_2$ is extremely important. In the current paper, we have studied the magnetic exchange interaction, Curie temperature in both Fe$_{16}$N$_2$ and V-doped Fe$_{16}$N$_2$.  The magnetic exchange constants were obtained by through first-principles based method and the Curie temperatures were calculated using both MFA as well as through performing MC simulation. In this paper, we also address the long standing question about the appearance of anomalously large magnetic moments. We organize our paper as follows: after a brief introduction about the computational methods in section II, in the following section (section III), we discuss the magnetic exchange interactions and Curie temperature, in the section IV we discuss about possible origin of the giant magnetic moments in the system. In the section V, we also introduce another possible application of this material,~i,e Fe$_{16}$N$_2$ might be used as a spin injector, beyond its application as permanent magnet. Finally, section VI presents our conclusions.
                             
\section{Computational method}
The electronic structure calculations were performed using first-principles methods within the frame-work of Density Functional Theory (DFT) with Perdew-Burke Ernzerhof exchange correlation energy functional\cite{pbe} based on a generalized gradient approximation. We used a projector augmented wave method as implemented in Vienna \textit{ab-initio} simulation package (\textsl{VASP})~\cite{vasp}.  Kohn-Sham wave functions of the valence electrons were expanded in plane wave basis with energy cut-off of 450 eV. The Brillouin zone sampling was carried out using Monkhorst Pack grid of 7x7x7 k-points. Ionic relaxation was performed using conjugate-gradient method, until forces were reduced to within 0.01 eV/Angstrom. The calculated structural parameters such as lattice constant, positions, magnetic moments etc. are then employed to obtain  the exchange parameters through  Korringa-Kohn-Rostoker
(KKR) multiple scattering theory. We have used Munich \textsl{SPR-KKR} package \cite{Ebert} to obtain the exchange parameters through a real space approach formulated by Liechtenstein \textit{et al.} which maps the total energy of the system to an effective Hamiltonian given by,
\begin{equation}
{\mathcal H}=-\sum_{i\neq j} J_{ij}{\bf m}_i{\bf m}_j
\end{equation}
Where ${\bf m}_i$ is the magnetic moment of i$^{th}$ site and $J_{ij}$ are the exchange constants.
The angular momentum expansion of the basis functions was taken up to l=3 . The k-space integration was performed using a grid of 280 k-points in the irreducible part of the Brillouin zone. We have used 30 complex energy points to perform the integration over the Green's function. The Curie temperature within MFA was calculated by solving the following coupled equations,
\begin{equation}
\langle {\bf m}^\alpha\rangle=\frac{2}{3K_BT}\sum_\beta J_0^{\alpha \beta} \langle {\bf m}^\beta \rangle
\end{equation}
where  $\langle {\bf m}^\alpha\rangle$ is the average z component of the magnetic moment in the sublattice $\alpha$. $J_0^{\alpha \beta}=\sum_{\bf r} J_{0{\bf r}}^{\alpha \beta}$. The summation in ${\bf r}$ is taken upto $|{\bf r}|/a=3.0$. To obtain accurate T$_C$, we performed Monte Carlo simulations with \textsl{VAMPIRE} atomistic spin dynamic program~\cite{vampire}. We used 67200 atoms with periodic boundary conditions. The DFT calculated magnetic moments (from \textsl{VASP}) and exchange constants (from \textsl{SPR-KKR}) were used as inputs. The Hamiltonian of the system contained exchange and anisotropy terms given by,
\begin{equation}
{\mathcal H}=-\sum_{ij}J_{ij}{\bf S}_i{\bf S}_j-K_u\sum_i({\bf S}_i\cdot{\bf e})^2
\end{equation}
Where ${\bf e}$ is the direction of easy axis and K$_u$ is the uniaxial anisotropy constant. The values of K$_u$ for pure and doped Fe$_{16}$N$_2$ were taken from our previous paper~\cite{KM}.
\section{Results and discussions}
Fe$_{16}$N$_2$ has a body center tetragonal structure with space group I$_4/mmm$ (number 139). Our PBE optimized lattice constants a=b=5.68\AA~, c=6.22\AA~ and position parameters, x$_{8h}$=0.243 and z$_{4e}$=0.294 are in good agreement with the ones measured by Jack \textit{et al.}~\cite{jack} (a=b=5.72\AA~ and c=6.29\AA~,x$_{8h}$=0.242,z$_{4e}$=0.293). For comparisons, we have considered other two variations of PBE, namely, PBE-sol~\cite{pbesol} and revised PBE \cite{rpbe}, which also yielded similar results. The average magnetic moment per Fe being respectively 2.43$\mu_B$ (PBE), 2.45 $\mu_B$ (PBE-sol) and 2.48 $\mu_B$ (rev PBE). As the change in magnetic moments are in the second decimal point we carried out other calculations with PBE functional only.

\subsection{Exchange constants and Curie temperature}
\begin{figure}
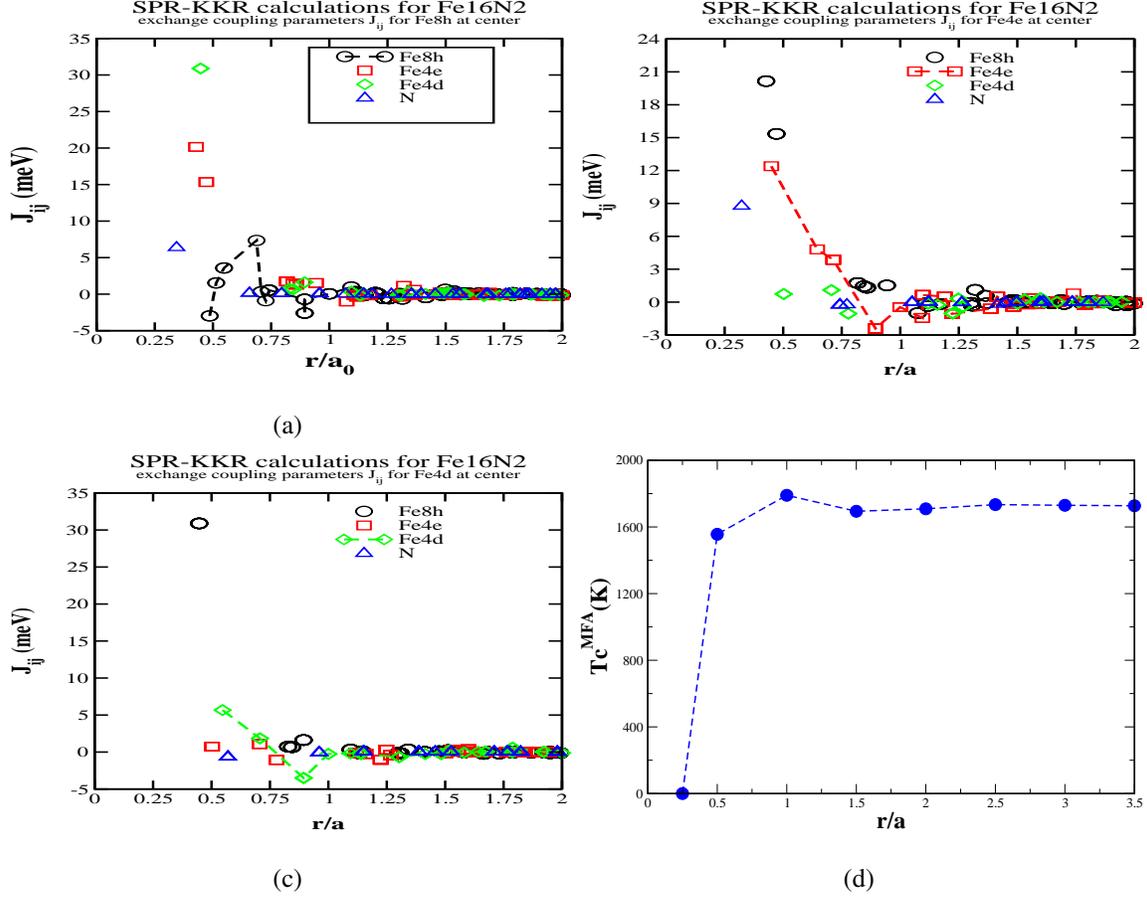

\subfigure[$ $  ]{\includegraphics[width=75mm,height=50mm]{1-1}}
\subfigure[$ $ ]{\includegraphics[width=75mm,height=50mm]{1-2}}
\subfigure[$ $  ]{\includegraphics[width=75mm,height=50mm]{1-3}}
\subfigure[$ $ ]{\includegraphics[width=75mm,height=50mm]{Tc}}
\caption{Exchange parameters for (a)Fe8h (b)Fe4e and (c)Fe4d as center. (d)The calculated Curie temperature for different values of radius of the sphere, within which all interactions were considered.}
\label{Exchange}
\end{figure}

\begin{figure}[!tbp]
  \centering
  \begin{minipage}[a]{0.6\textwidth}
    \includegraphics[width=\textwidth]{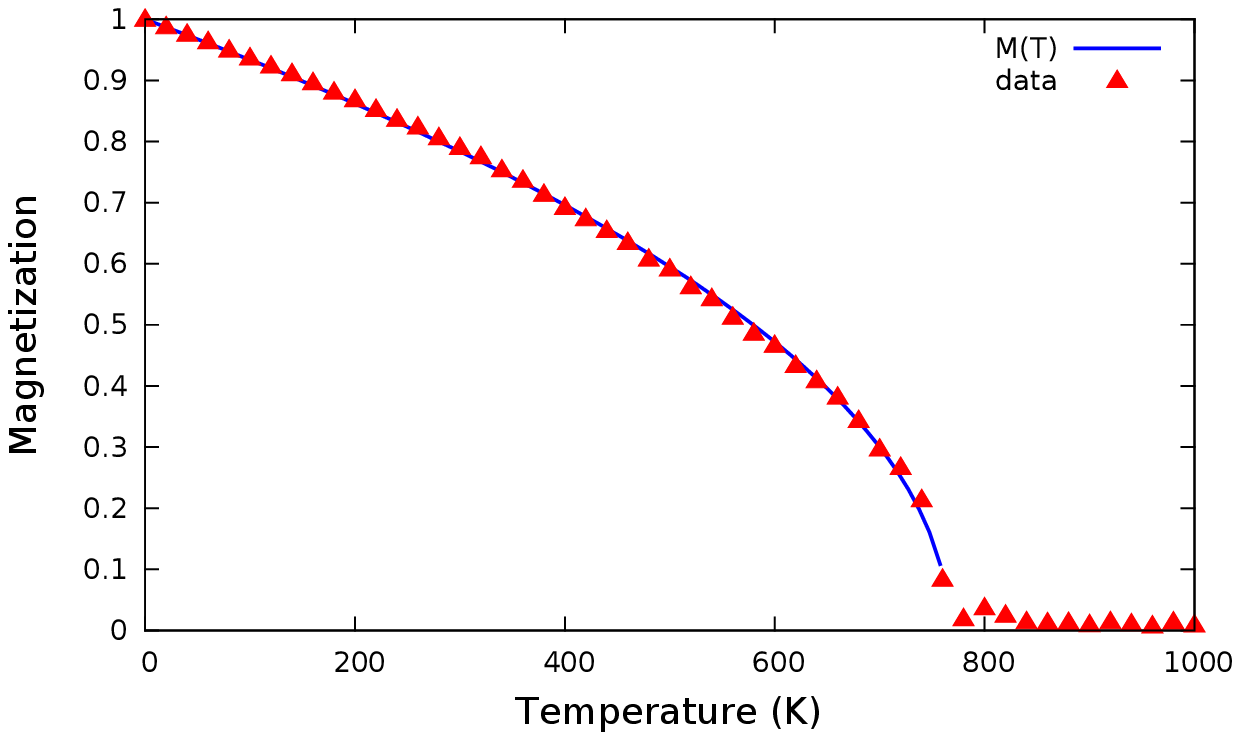}
     
  \end{minipage}
  \hfill
  \begin{minipage}[b]{0.6\textwidth}
    \includegraphics[width=\textwidth]{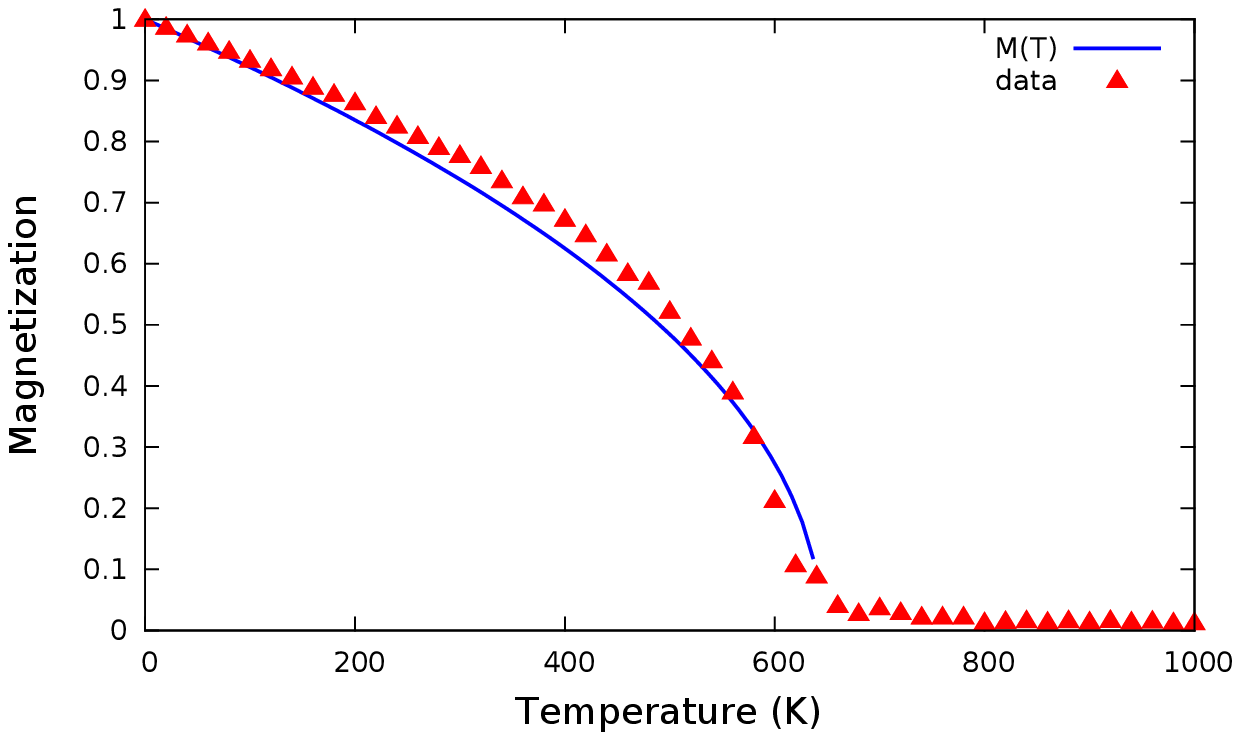}
  \end{minipage}
  \caption{Monte Carlo simulation of temperature dependent magnetization. Top panel: for pure Fe$_{16}$N$_2$, bottom panel: V-doped Fe$_{16}$N$_2$ }
  \label{TC-MC}
\end{figure}

In the Fig.\ref{Exchange}, we show the calculated exchange constants obtained with SPR-KKR method with PBE functional. Though the trend of the results have some similarities to that of Ke \textit{et al.}~\cite{kirill} obtained with TB-LMTO method within local density approximation (LDA) and GW method, yet significant differences are present. The largest among all interactions is between between 8h-4d sites (30.89 meV) at distance 2.54\AA as shown in Fig.1(a)), while for Ke \textit{et al.} it was either 4e-4e (within atomic sphere approximation) or 8h-4e (within fullpotential method). The next strongest interaction from our calculation is among 8h-4e sites (20.15 meV) (Fig.1(a)) at a distance of 2.43~\AA between two octahedral Fe atoms. The third strongest interaction (15.34 meV) is between octahedral 8h and 4e sites at distance 2.67~\AA. From the structural consideration, this interaction should be mediated through 90$^\circ$ Fe8h-N-Fe4e superexchange interaction. Among inter sublattice interaction 4d-4e interaction is weakest (0.74 meV) (Fig.1(b)). From Fig.1(a) it is clear the 8h-8h interaction antiferromagnetic (-2.97 meV) at smallest distance (2.76 \AA~). This interaction corresponds to 90$^\circ$ Fe8h-N-Fe8h superexchange interaction. The strongest Fe8h-Fe8h interaction is ferromagnetic (3.58 meV) at 3.9~\AA~ and it is due to 180$^\circ$ Fe8h-N-Fe8h superexchange interaction. Among the intra-sublattice interactions, 4e-4e interaction is strongest (12.39 meV), which corresponds the interaction between two Fe4e atoms in neighbouring octahedra. The next strongest 4e-4e interaction is between two epical Fe atoms in the Fe-N octahedra an is therefore mediated through 180$^\circ$ superexchange interaction. The strongest 4d-4d interaction is 5.3 meV at a pair distance of 3.1\AA.\\
In the Fig.\ref{Exchange} (d), we show the variation of mean field Curie temperature as a function of the cluster radius $|{\bf r}|$. It can be seen that even for a small radius of about $|{\bf r}|=0.5 \AA$ , the T$_C$ is much higher than the experimental value (813 K), which shows that mean field picture can not explain the T$_C$ of this material. It is well-known that MFA overestimates the T$_C$ due to neglect of the thermal fluctuation of magnetization. However, in the present case, MFA overestimates T$_C$ by almost 50\%, which is quite large. Such overestimation of T$_C$ of Fe$_{16}$N$_2$ within MFA was observed by Ke \textit{et al.} also, however they attributed this to the lack of proper measurement of experimental  T$_C$. We, on the other hand, point out that such consistent huge overestimation of T$_C$ within MFA reflects a low dimensional behaviour of the spin system rather than a pure 3D one.~The origin of such quasi-2D behaviour could be due to the poor connectivity between certain spins in different directions. There is almost non-existent  exchange between spins at the 4e and 4d sites and a very weak 4d-4d interaction as can be seen from the Fig.\ref{Exchange}.\\
To obtain better results, we performed MC simulation using Metropolis method~\cite{metro}. In the Fig\ref{TC-MC}, we show our results for the MC simulations. In order to obtain the T$_C$, we fit the temperature dependent magnetization data to the function $M(T)=(1-\frac{T}{T_C})^\beta$, which yields a T$_C$ of about 765 K Fig.\ref{TC-MC}(a) which is much close to the experimental compared to the MFA results obtained by us well as by Ke \textit{et al.}. For a comparison, we also performed MC simulation for the V doped Fe$_{16}$N$_2$. As it was demonstrated in our previous study~\cite{KM} that, Fe$_{16}$N$_2$ can be stabilized by doping V at 8h site, we performed the MC calculations with exchange constants obtained with the SPR-KKR method by replacing one Fe-8h with V. The Fig.\ref{TC-MC}(b) shows the obtained results. The T$_C$ evaluated after fitting is about 640 K, which is somewhat less than the bulk Fe$_{16}$N$_2$, but still above the room temperature.

\section{u\lowercase{nderstanding the \textit{giant magnetic} moment}}
In this section, we discuss about the origin of the so called giant magnetic moment. As already mentioned, in the introduction that, the experimental findings about such moments are not consistent. Therefore the reliability of such reports are not completely doubtless. However in the following we try to analyse the possibility of any \textit{giant} moment in this system. We discuss here the following two factors which could influence the saturation magnetization: (1) role of structure, which we study through using constrained magnetic calculation,
(2) existence of localized or semi-localized d-states, which we justify from the magnetic susceptibility susceptibility and the nature of the exchange interactions.

\subsection{Structural sensitivity and magnetism: constrained magnetic calculations}
To understand the relation between the structure and magnetism we performed constrained moment calculation with PBE exchange correlation function. The magnetic moment of the cell was fixed such a way that each Fe site has a magnetic moment of about 3$\mu_B$. With such constrain, we relaxed the cell geometry completely, which gave lattice parameters a=b=5.84\AA and c=6.34\AA. It is interesting to notice that c/a ratio remains close 1.10 (as in the unconstrained case) showing that large magnetic moment does not bring any additional structural anisotropy such as additional teragonality etc. The space-group (No. 139) remains same after the structural optimization with giant moments are introduced. The only  difference appears in the position parameters with z$_{4e}$=0.3, and x$_{8h}$= 0.240. This results slight increase in the distance between N and Fe4e and Fe8h irons. Next, we performed a calculation which involves optimization of positions but the volume of the cell is fixed to the one which was obtained through the  above mentioned constrained magnetic calculation. We allowed the atomic spin moments to change with the electronic steps in this case. This resulted a magnetic moment of 2.52$\mu_B$ per Fe atom just slightly bigger than the values reported in Table I. This calculation suggests a weak correlation between the structure  and magnetic moment in Fe$_{16}$N$_2$. It is therefore very unlikely that such large moments may appear due to some sort of strain effects arising from substrates or other means. The origin of giant magnetic moment should be related to some missing physics such as partial localization of d-electrons in the Fe-N octahedra as discussed below. Such partial localization of d-electrons was discussed by Ji \textit{et al.}~\cite{myst}. However they demonstrated it in terms of the difference in charge density between inside and outside the Fe$_6$N octahedra.
 
\subsection{Partial localization of d-electron in Fe$_6$N octahedra}
\begin{figure}
\centering
\includegraphics[width=130mm,height=120mm]{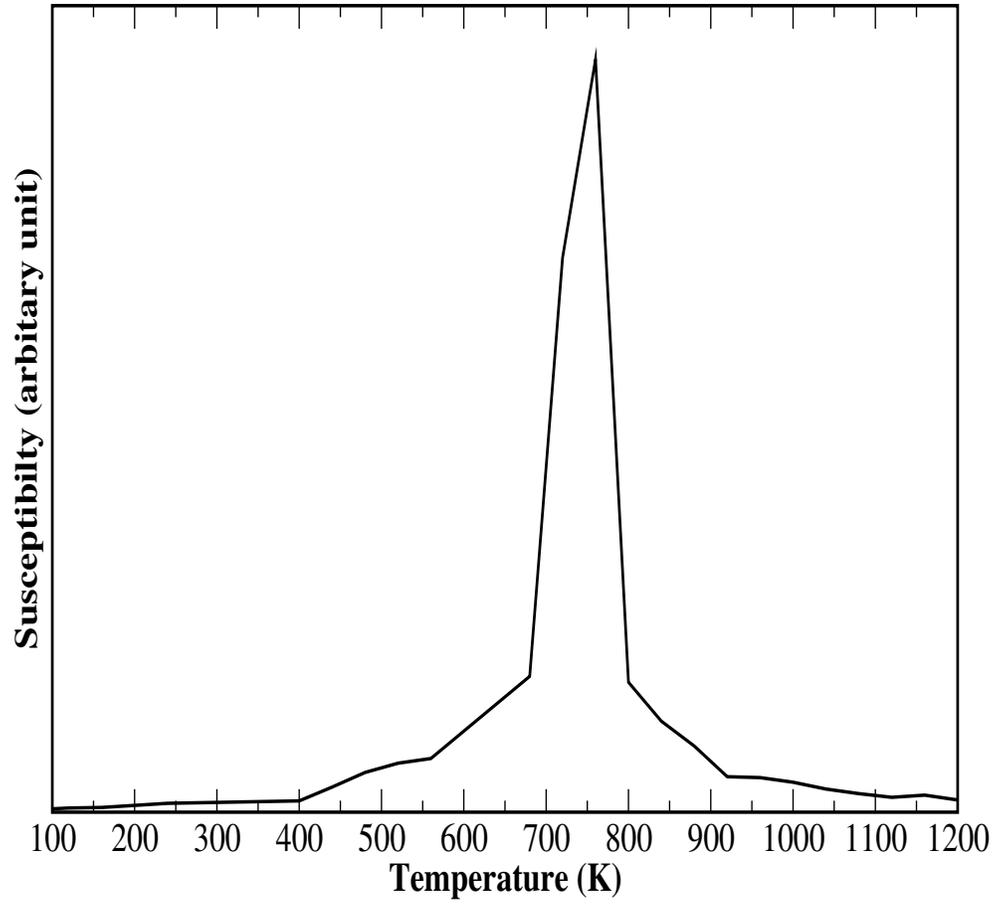}
\caption{Magnetic susceptibility of bulk Fe$_{16}$N$_2$ obtained from MC simulations}
\label{susc}
\end{figure}

\begin{figure}[H]
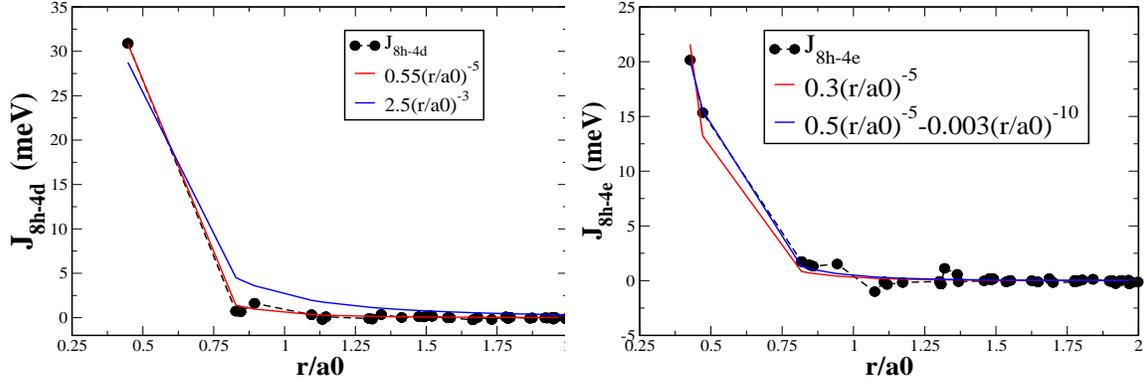
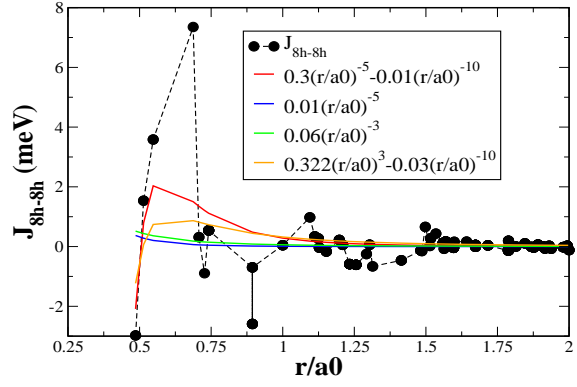

\subfigure[$ $  ]{\includegraphics[width=75mm,height=50mm]{8h-4d}}
\subfigure[$ $ ]{\includegraphics[width=75mm,height=50mm]{8h-4e}}
\centering\subfigure[$ $  ]{\includegraphics[width=75mm,height=50mm]{8h-8h}}
\caption{The numerically obtained magnetic exchange interactions involving the octahedral 8h site (a)8h-4d (b)8h-4e (c)8h-8h interactions. These values are fitted with different models as discussed in the section IV.}
\label{interactions}
\end{figure}

In the Fig.\ref{susc}, we show the calculated magnetic susceptibility as function of temperature. The susceptibility shows a sharp peak at T=T$_C$, which indicates the presence of localized states in this material. It should be noted that for itinerant systems the magnetic susceptibility usually shows a broad peak below the T=T$_C$. It is therefore meaningful to investigate the possibility of localized electronic states which contribute to the magnetic moments. It can be seen that interactions involving octahedral 8h sites shows strong tendency towards superexchange interaction. In the Fig.\ref{interactions}, we analyse the magnetic exchange interactions (obtained through SPR-KKR) involving 8h sites within the framework of the tight binding model. Within the tight-binding model, the intersite hopping integral, $t_{dd}$,  between the Fe sites involving only d-orbitals can be expresses as $t_{dd}\sim {(r/a0)}^{-5}$~\cite{Hari}. In the Fig.\ref{interactions} (a), for the case of 8h-4d interactions we fit the exchange constants with two itinerant types interactions, namely double exchange and RKKY type. Typically, the double exchange interaction is directly proportional to  $t_{dd}$ i,e ${r/a0}^{-5}$, while the RKKY interaction proportional to ${(r/a0)}^{-3}$. From the figure it can be seen that $J_{8h-4d}$ fit extremely  well with double exchange type. Due its double exchange nature, 
$J_{8h-4d}$ interaction is strongly ferromagnetic. For the 8h-4e interactions, however we can see (Fig.\ref{interactions} (b)) that a linear combination of double exchange and superexchange interaction of the form $J=\alpha{(r/a0)}^{-5}+\beta{(r/a0)}^{-10}$ works well (as the superexchange interaction has $t_{dd}^2$ dependence). $\alpha$ and $\beta$ being the fitting parameters. Finally, for 8h-8h interactions, we fit the exchange constants with four types of interactions, (1) linear combinations of double exchange and superexchange (red curve) (2) double exchange (blue curve) (3) RKKY (green curve) and (4) combination of RKKY and superexchange. From the figure, it can be understood that, the behaviour of the data has strongest resemblance with the case where fitting is done with the combination of double exchange and superexchange interactions. The substantial tendency of the exchange interactions  within the octahedra towards superexchange interactions therefore suggests 
about partial localization of electron states at 8h-sites even though not very strong.

Note that among itinerant interactions, it is the double exchange one which is the strongest in this material. This was first predicted by Sakuma \textit{et al.}~\citep{sakuma}. Our conclusion about the high moment is as follows: due to the presence of N, there is a moderate correlation effect in the octahedral region which localizes partially the d-states within this region (especially the 8h states) while a strong tendency towards double exchange interaction suggests that there is a reduction of the exchange splitting of d-electrons which adds further enhancement of the moments. 

\begin{figure}
\centering
\includegraphics[width=130mm,height=120mm]{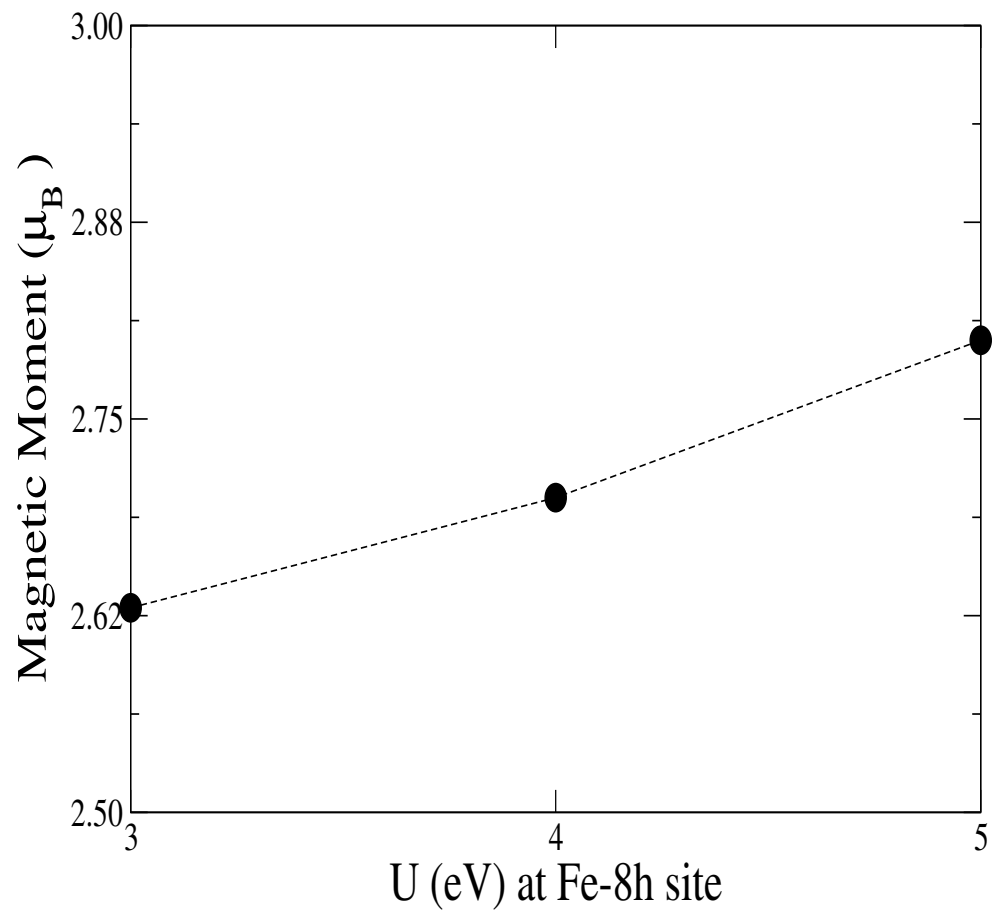}
\caption{Magetic moment per Fe for the different U values at 8h site}
\label{8hU}
\end{figure}

To understand at what extent this  partial localization of the octahedral d-states may effect the magnetic behaviour or particularly the saturation moment of Fe$_{16}$N$_2$, we computed the magnetic moments per unit cell using DFT+U method~\cite{dftu} with different values of U as shown in the Fig.\ref{8hU}. It can be seen that switching on localization effect in a moderate level (3-5 eV) gives rise to magnetic moment as high as $\sim 2.8\mu_B$. It should be noted that for typical Fe-based insulating systems the usual choice for U is about 6 eV. As Fe$_{16}$N$_2$ is a metal, the U value should be smaller than 6 eV. 

\section{F\lowercase{e$_{16}$}N$_2$ \lowercase{as a spin injector}}
\begin{figure}
\includegraphics[width=135mm,height=130mm]{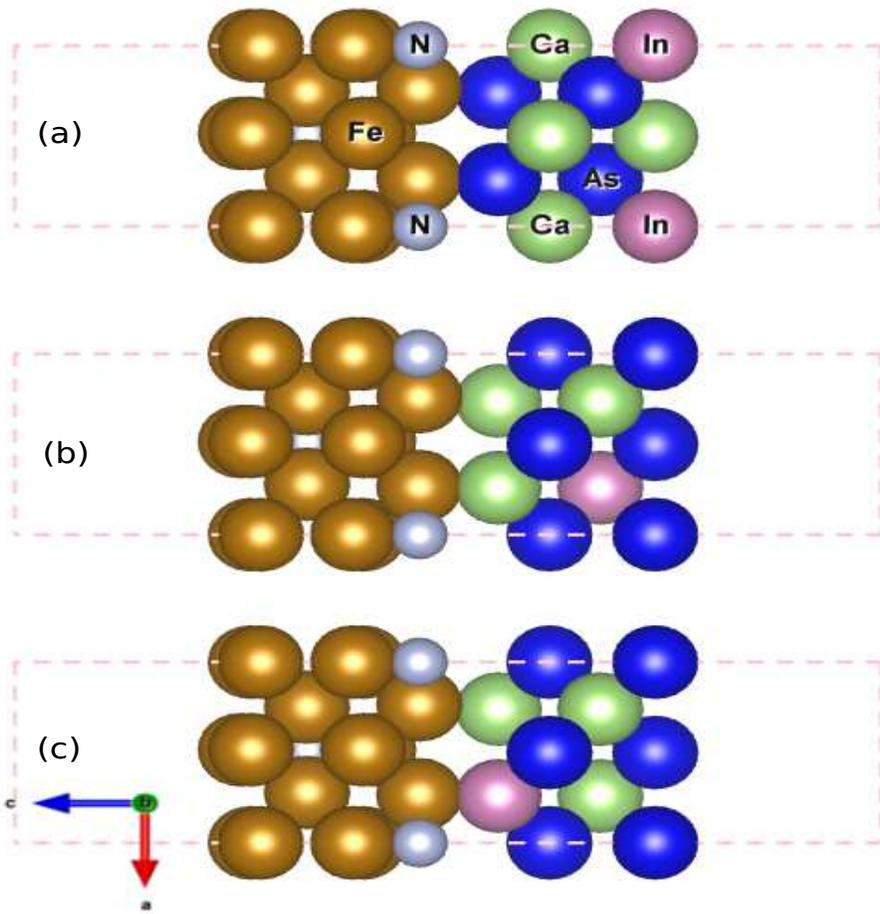}
\caption{Fe$_{16}$N$_2$(001)/InGaAs(001) thin with 4-monolayers of metal and semiconductor for three configurations.}
\label{structures1}
\end{figure}

\begin{figure}
\centering
\includegraphics[width=130mm,height=120mm]{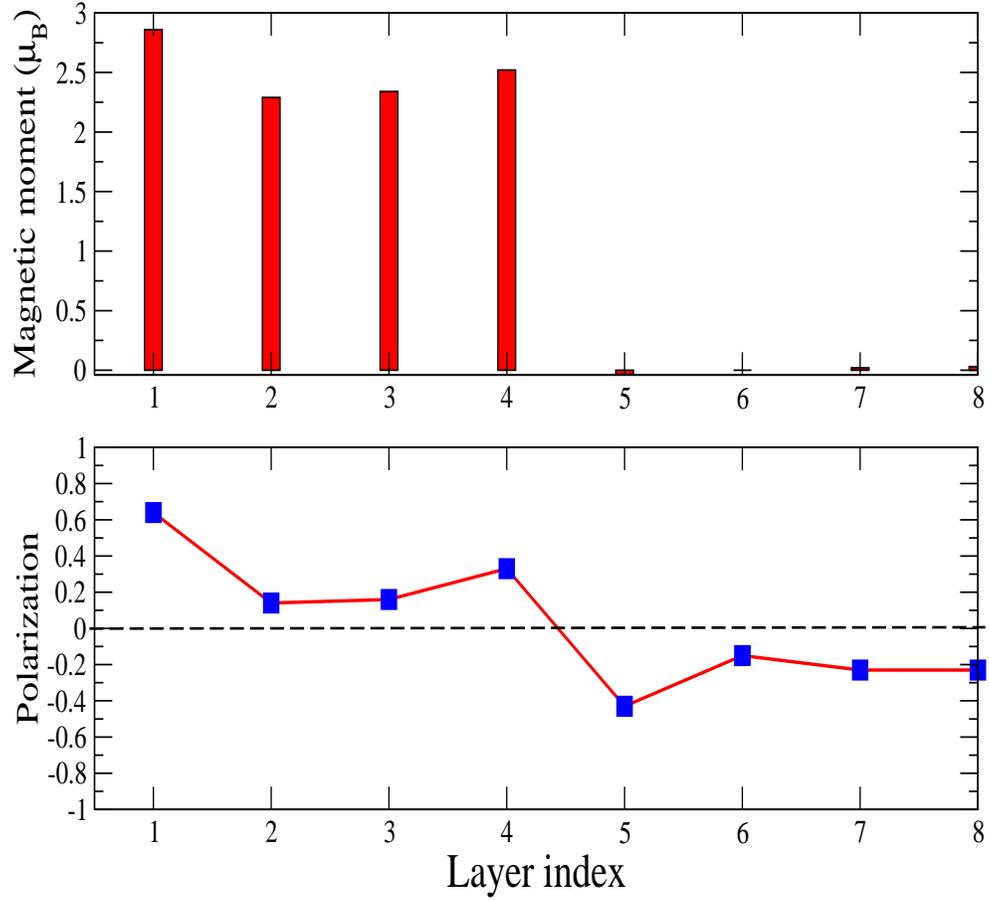}
\caption{Layer resolved magnetic moments/per atom for the Fe$_{16}$N$_2$(001)/InGaAs(001) thin film configuration(b) of FIG.\ref{structures1}.}
\label{polarization}
\end{figure}

The bulk spin polarization of the PBE-optimized  Fe$_{16}$N$_2$ we obtained is about 87\%, which is quite large compared to the elemental metallic ferromagnets such as Fe (45\%), Co (42\%) and Ni (27\%) \cite{sp}. Such large spin polarization is almost comparable to the half-metallic Heusler alloys ($\sim$ 100\%) which were proposed to be the best spin-injectors for the semiconductors~\cite{best}. However, till the date, the Heusler alloys were not very successful spin injectors due to their large lattice mismatch with the semiconductors~\cite{poor1,poor2}. It is well-known than Fe$_{16}$N$_2$ can be easily grown on III-V semiconductors such as GaAs or InGaAs (historically the so called giant moment was observed in epitaxially grown single crystals of Fe$_{16}$N$_2$ either on  InGaAs (001)or GaAs(001) film). As Fe$_{16}$N$_2$ has both large spin polarization as well as it has good lattice matching with semiconductors we looked at its another important possible application, viz, its application as a spin injector as follows. 
                 We performed calculations with 4 monolayers (ML) of Fe$_{16}$N$_2$ (001)on 4 ML of InGaAs (001) as well as 8 ML of InGaAs (001). The lattice parameters of bulk InGaAs is obtained first. The optimized structure was used to construct the metal/semiconductor (001) interfaces as shown in the Fig.\ref{structures1}. The magnetic moments per Fe for configurations (a), (b) and (c) are respectively 2.48$\mu_B$,~2.51$\mu_B$ and 2.52$\mu_B$ respectively. It can be seen that the the magnitude of the moment depends on the nature of the interface. The magnetic moment is higher for the interface containing either Ga or In. These moments are certainly not \textit{giant}, however comparable to  the maximum value of the saturation moment predicted by so called Slater-Pauling curve~\cite{slaterpauling}. Then we look at the possibility of spin injection from ferromagnet (FM) to semiconductor (SC) in such systems. As an example, we have computed the layer resolved spin polarization $P_L=\frac{D^\uparrow_L(E_F)-D^\downarrow_L(E_F)}{D^\uparrow_L(E_F)+D^\downarrow_L(E_F)}$, where L is the layer index $D^{\uparrow(\downarrow)}_L(E_F)$ is the density of states of states at the Fermi energy for the majority (minority) spin carriers respectively for the interface (b) described in the Fig.\ref{structures1}. The results are shown in the Fig.\ref{polarization} which shows layer dependent magnetic moment in the top panels. In the Fe$_{16}$N$_2$ side, all the four layers are having moments close to 2.5$\mu_B$ (as the average is around 2.51$\mu_B$) while the spin polarization (bottom panel) reduces as we approach to the interface. In the semiconductor side on the other hand (5$^{th}$ to 8$^{th}$ layer) spin polarization is quite high and is non-negligible even upto the last layer. If we consider the average spin injection efficiency it is about 30\% averaged over all the four layers which is quite higher than(almost double) than what is observed for Fe/GaAs heterostructures~\cite{fe-h}. This implies a long spin diffusion length in Fe$_{16}$N$_{2}$/InGaAs systems. It is interesting to observe that the spin polarization changes sign across the interface which means the spin imbalance in the semiconductor side is due the minority spin.\\
\indent Fe$_{16}$N$_2$ seems to be very good spin injector for III-V semiconducting systems like InGaAs. As shown in our previous study~\cite{KM}, small V doping stabilizes Fe$_{16}$N$_2$, its interface with III-V semiconductor might be important for spin-optoelectronics or spintronic~\cite{spintronic1,spintronic2} applications. 
\section{Conlusions}
We have studied the exchange interactions and magnetic properties of Fe$_{16}$N$_2$ and V-doped Fe$_{16}$N$_2$ using first principles based methods. The magnetism and T$_C$ are dominated by the exchange interactions between Wyckoff sites of different types such as 8h-4d and 8h-4e. Though for 8h-4d and 8h-4e the and first nearest neighbours are of itinerant in nature, our calculations suggest that remaining important interactions involving 8h sites (between 8h-4e and 8h-8h) are of superexchange type and due to the localized d-states. This suggest that d-electrons with Fe$_6$N octahedra exhibit both itinerant and localized behaviour and the origin of the large saturation moment could be due to such dichotomy. We also pointed out that in addition to the permanent magnet application, Fe$_{16}$N$_2$ can also find important spintronic applications due to its high spin polarization and fantastic lattice matching with the III-V semiconductors.
\section{Acknowledgement and Funding}
This work was supported by NRF grant funded by MSIP, Korea(No. 2009-0082471 and No. 2014R1A2A2A04003865),the Convergence Agenda Program (CAP) of the Korea Research Council of Fundamental Science and Technology (KRCF) and GKP (Global Knowledge Platform) project of the Ministry of Science, ICT and Future Planning.

\newpage
\begin{table}
\begin{tabular}{|c|c|c|c|c|c|c|c|}
\hline
GGA & a (\AA) & c (\AA) & m$_{8h}$ ($\mu_B$) & m$_{4e}$ ($\mu_B$) & m$_{4d}$ ($\mu_B$) &x$_{8h}$ & z$_{4e}$ \\
\hline
\hline
PBE & 5.68 & 6.22 & 2.36 & 2.18 &2.82 & 0.242 & 0.293\\
PBE-sol & 5.75& 6.28 & 2.40 & 2.17 &2.84 &0.242 & 0.294\\
Revised PBE & 5.75 & 6.28 & 2.42 & 2.22 &2.86 &0.243 &0.294\\
\hline
\end{tabular}
\caption{Lattice constants, magnetic moments in three different types of GGA calculations.}
\end{table}
\clearpage
\newpage
\clearpage
\newpage
\clearpage
\newpage


\newpage
\clearpage

\clearpage

\newpage
\clearpage

\newpage
\clearpage

\newpage
\clearpage

\newpage
\begin{thebibliography}{99}
\bibitem{giant1}T. K. Kim and M. Takahashi, \textit{New Magnetic Material Having Ultrahigh Magnetic Moment}~Appl. Phys. Lett., {\bf 20}, 492 (1972)
\bibitem{giant2}Y. Sugita, K. Mitsuoka, M. Komuro, H. Hoshiya, Y. Kozono, and
M. Hanazono, \textit{Magnetism of $\alpha^"$‐Fe$_{16}$N$_2$ (invited) } J. Appl. Phys. {\bf 70}, 5977 (1991)
\bibitem{giant3} J. M. D. Coey,\textit{The magnetization of bulk $\alpha^"$ Fe$_{16}$N$_2$}  J. Appl. Phys. {\bf 76}, 6632 (1994)
\bibitem{jack}K. H. Jack, \textit{The iron-nitrogen system: the preparation and the crystal structures of nitrogen-austenite($\gamma$) and nitrogen-martensite($\alpha$')}~Proc. R. Soc. A, {\bf 208}, 200 (1951)
\bibitem{confirmed1} M. Takahashi and H. Shoji, \textit{$\alpha^"$-Fe$_{16}$N$_2$ problem — giant magnetic moment or not?}~J. Magn. Magn. Mater. {\bf 208}, 145 (2000)
\bibitem{confirmed2}J. M. Cadogan, \textit{Are there Giant Magnetic Moments in Fe-nitrides?} Aust. J. Phys. {\bf 50}, 1093 (1997)
\bibitem{failed}Coey J M D, O’Donnell K, Qinian Q, Touchais E and Jack K H, \textit{
Journal of Physics: Condensed Matter
The magnetization of alpha"Fe$_{16}$N$_2$} J. Phys. Condens. Matter {\bf 6} L23, 1994
\bibitem{myst}N. Ji, X. Liu, and J.-P. Wang, \textit{Theory of giant saturation magnetization in $\alpha^"$-Fe$_{16}$N$_2$} New J. Phys. {\bf 12}, 063032 (2010)
\bibitem{kirill}Liqin Ke, Kirill D. Belashchenko, Mark van Schilfgaarde, Takao Kotani, and Vladimir P. Antropov, \textit{Effects of alloying and strain on the magnetic properties of Fe$_{16}$N$_2$} Phys. Rev. B, {\bf 88}, 024404 (2013)
\bibitem{KM} Krishnamohan \textit{et al.} \textit{Improving the Stability and Magnetic Hardening of Fe$_{16}$N$_2$ by Alloying:
A First-principles Study}(Unpublished)
\bibitem{metro}K. Binder, \textit{Applications of Monte Carlo methods to statistical physics } Rep. Prog. Phys. {\bf 60}, 487 (1997)
\bibitem{pbe}Perdew,~J.P, Burke,~K., Ernzerhof,~M., \textit{Generalized gradient approximation made simple}~Physical Review Letters,{\bf 77}, 3865 (1996)
\bibitem{vasp}Kresse,~G.,~Furthm\"uller,~J., \textit{Efficient iterative schemes for ab initio total-energy calculations using a plane-wave basis set
}~Physical Review B, {\bf 54}, 11169 (1996)
\bibitem{Liechtenstein}Liechtenstein A I, Katsnelson M I, Antropov V P and
Gubanov V A,~\textit{Local spin density functional approach to the theory of exchange interactions in ferromagnetic metals and alloys}, J. Magn. Magn. Mater. {\bf 67} 65 (1987)
\bibitem{Ebert}Ebert, H Fully relativistic band structure calculations
for magnetic solids: formalism and application Electronic Structure and Physical Properties of Solids (Lecture Notes in Physics vol 535) ed H Dreyss (Berlin: Springer) p 191 (2000)
\bibitem{vampire}Evans R F L, Fan W J, Chureemart P, Ostler T A, Ellis M O A
and Chantrell R W, \textit{Atomistic spin model simulations of magnetic nanomaterials},  J. Phys.: Condens. Matter, {\bf 26} 103202 (2014)
\bibitem{pbesol} J. P. Perdew, A. Ruzsinszky, G. I. Csonka, O. A. Vydrov, G. E. Scuseria, L. A. Constantin, X. Zhou, and K. Burke, \textit{Restoring the Density-Gradient Expansion for Exchange in Solids and Surfaces},~Phys. Rev. Lett. {\bf 100}, 136406 (2008)
\bibitem{rpbe} B. Hammer, L. B. Hansen and J. K. N{\o}rskov, \textit{Improved adsorption energetics within density-functional theory using revised Perdew-Burke-Ernzerhof functionals},~ Phys. Rev. B {\bf 59}, 7413 (1999)
\bibitem{dftu}S. L. Dudarev, G. A. Botton, S. Y. Savrasov, C. J. Humphreys and A. P. Sutton, \textit{Electron-energy-loss spectra and the structural stability of nickel oxide:~An LSDA+U study}, Phys. Rev. B {\bf 57}, 1505 (1998)
\bibitem{Hari}W. A. Harrison,\textit{Electronic structure and the properties of solids}, Dover publications, 1988
\bibitem{sakuma} A. Sakuma, \textit{Electronic and magnetic structure iron nitride},~ J. Appl. Phys., {\bf 79}, 5570 (1996)
\bibitem{sp}R. J. Soulen, \textit{Measuring the Spin Polarization of a Metal with a Superconducting Point Contact}, Science, {\bf 282},  85, (1998)
\bibitem{best}R. Farshchi and M. Ramsteiner, \textit{Spin injection from Heusler alloys into semiconductors: A materials perspective}, J. Appl. Phys. {\bf 113}, 191101 (2013)
\bibitem{poor1}M. Hashimoto, J. Herfort, H.-P. Schonherr, and K. H. Ploog, \textit{Epitaxial Heusler alloy Co2FeSi∕GaAs(001)
hybrid structures}, Appl. Phys.Lett. {\bf 87}, 102506 (2005)
\bibitem{poor2}M. Hashimoto, J. Herfort, A. Trampert, H.-P. Schonherr, and K. H.
Ploog, \textit{Growth temperature dependent interfacial reaction of Heusler-alloy Co2FeSi/GaAs(001) hybrid structures}, J. Phys. D: Appl. Phys. {\bf 40}, 1631 (2007)
\bibitem{slaterpauling}P. Mohn, Magnetism in the solid state: an introduction, Vol. 134, Springer Science \& Business Media, (2006)
\bibitem{fe-h}T.H. Lee \textit{et. al},\textit{Temperature dependence of spin injection efficiency in an epitaxially grown Fe/GaAs hybrid structure},  Journal of Magnetism and Magnetic Materials, {\bf 321}, 3795-3798 (2009)
\bibitem{spintronic1} S. Bhattacharjee, S. Singh, D Wang, M Viret and L Bellaiche, \textit{Prediction of novel interface-driven spintronic effects}, Journal of Physics: Condensed Matter, {\bf 26},315008 (2014)
\bibitem{spintronic2} D. D Awschalom and M. E. Flatte, \textit{Challenges for semiconductor spintronics}, Nature Physics, {\bf 3}, pages 153–159 (2007)

\end{thebibliography}
\end{document}